
\input amstex
\documentstyle{amsppt}
\document
\NoRunningHeads
\NoBlackBoxes
\font\ninebf=cmbx9
\font\nineit=cmti9
\font\special=cmr6
\topmatter
\title
Remarks on deformed and undeformed \\ Knizhnik-Zamolodchikov
equations.
\endtitle
\author
Fedor A. Smirnov\\
{\special RIMS,  Kyoto  University, Kyoto,  606,  JAPAN}\\
{\special and}\\
{\special Steklov Mathematical Institute,
  St. Petersburg 191011,  RUSSIA}
\endauthor
\abstract
Deformed and undeformed KZ equations are considered for
$k=0$. It is shown that they allow the same number of solutions,
one being the asymptotics of others. Essential difference in analitical
properties of the solutions is explained.
\endabstract
\endtopmatter

In this paper certain beautiful mathematical structures will
be explained which arise when studying the deformations
of famous Knizhnik-Zamolodchikov equations [1].
These deformations were proposed in [2]. We shall be considering
the case of zero central extension for which the deformed
equations coincide with Form Factors Equation [3] as it
has been explained in [4]. We shall show that there exists
exactly the same number of solutions to deformed equations
as to usual ones although the properties of the solutions
are quite different.

The stress will be done on the braiding properties of the
solutions. We shall show that the deformation leads to
trivialization of braiding. That is opposite to the usual
belief that the deformation should lead to more complicated
braiding. The trivialization of braiding is responsible
for the application of the deformed equations to the theory
of particles, i.e. to the description of form factors.
Also, it is closely connected with certain ``quantization''
of Riemann surfaces which will be discussed below.

{\bf Acknowledgement} This paper presents a talk given at Mathematical
Department of Nagoya University. I am grateful to K. Aomoto and A. Tsuchiya for
the kind invitation to give this talk. Special thanks are due to
T. Nakanishi for his warm hospitality. The work was done under JSPS
Postdoctoral Fellowship for Foreign Researches.

1. {\bf Knizhnik-Zamolodchikov equations for $k=0$.}

Knizhnik-Zamolodchikov equations for the affine algebra
$\widehat{sl}(N)$
can be written as follows:
\define\la#1{\lambda_{#1}}
\define\laaa#1{\lambda_1 ,\cdots , \lambda_{#1}}
$$ \bigl( (\frac{k+N}{2})\frac{d}{d\la i} +
 \sum_{j\neq i}\hat{r}_{i,j} (\la i - \la j )\bigr)
f(\laaa m)=0$$
where $k$ is the central charge, a finite-dimensional
representation of $sl(N)$ is associated to every $\la i$,
$\hat{r}_{i,j}$ is classical r-matrix acting in the tensor
product of i-th and j-th representations:
$$ \hat{r}_{i,j} (\la i - \la j)=\frac{K^{ab}t^a_i \otimes t^b_j}
{\la i -\la j},\tag 1$$
$K$ is Killing form, $ t^a_i,t^b_j $ are generators of $sl(N)$
acting in corresponding representations. The function $f$ belongs
to the tensor product of the spaces of the representations.

We shall be considering the following particular case of
the equations: the algebra under consideration will be $sl(2)$,
we shall deal only with fundamental two-dimensional representations,
we take rather unusual value of the central charge: $k=0$. Also we change
the definition of $r_{i,j}$ to
$$r_{i,j}(\la i-\la j)={1\over 2}\frac{P_{i,j}}{\la i - \la j}\tag 2$$
($P_{i,j}$ is permutation) which differs in our case from (1) by
${1 \over 4} I$. With that change of r-matrix some irrelevant multiplier
($ \prod_{i<j}(\la i -\la j)^{1/4}$) is removed from the solutions.
We restrict ourselves with those solutions which are singlets
(belong to one-dimensional representation) under the global $sl(2)$.
So, we deal with the equations
$$\bigl( \frac{d}{d\la i}+ \sum_{i\neq j}r_{i,j}
(\la i - \la j)\bigr)f(\laaa {2n})=0 \tag 3$$
with $r_{i,j}$ given by (2).

As it has been mentioned the consideration of $k=0$ is rather
unusual, so we have to comment on it. Usually $k\in Z_+$ are
considered which are responsible for the applications to the
Conformal Field Theory, $\la i$ then are points in the
coordinate space. One can try to apply the deformed KZ equations
which are to appear soon to certain lattice deformation of
WZNW model ( may be to the one described in [5] ). In that case,
certainly, $k\in Z_+$ should be considered for the deformed
equations either. However, the applications of the deformed
equations we are interested in are completely different:
$\la i$ for us will belong rather to the momentum space
describing the rapidities of physical particles. There are good reasons
for $k$ to equal zero in that case. One can ask about the meaning
of undeformed KZ equations in this situation. It is a
complicated question the answer to which is not quite clear now.
The following can be said, however. The classical limit of the
deformed equations should describe not the ultraviolet
limit in a sense of perturbations of CFT [6], but rather certain
quasiclassical limit for asymptotically free field theories
which is connected with periodical problem for classical
integrable equations. This understanding follows from the
logic of the paper [4]. It should be emphasized that the periodical
problem is used not in usual fashion as regularization of space
interval, but rather as that associated with monodromy
around one point in the space-time. So, in certain sense the situation is
close to that considered in [7] where the finite-dimensional
quantum group responsible for the quantization of WZNW model
arises from the monodromy of classical equation. In our case
the infinite-dimensional quantum group (double of Yangian [8,11,4])
responsible for the quantization of asymptotically free
massive field theory [9,10,11,4] should arise from the monodromy
of classical integrable equation of KDV type.
In the $sl(2)$ case the periodical problem in
question is closely connected with hyper-elliptic surfaces (HES) which
parametrize the solutions.
So, no wonder that the solutions of (3) are closely related to
HES. Continuing these reasonings we would like to identify the particles
in quantum theory with the moduli of classical periodical problem.
But still much work remains to be done in this direction.

Anyway, we consider $k=0$. So, let us write down the solutions to
the equations (3). Using the usual basis in two-dimensional
space:
$$e^+=\binom 10 ,\  e^-=\binom 01$$
we shall denote the components of $f(\laaa {2n})$ in the natural
basis of the tensor product
\define\eee#1{\epsilon_1, \cdots, \epsilon_{#1}}
$$ e^{\eee {2n}}=e^{\epsilon_1}\otimes\cdots\otimes e^{\epsilon_{2n}}\tag 4$$
by $f(\laaa {2n})_{\eee {2n}}$. We are looking for the singlet
solutions of the equations which means in particular that
$\sum \epsilon_i=0$. For each particular component
$f(\laaa {2n})_{\eee {2n}}$ the choice of $\eee {2n}$
induces a partition of $\Lambda=$\linebreak $\{\laaa {2n}\}$ into
$\Lambda^+ =\{ \la j : \epsilon _j =+\}$ and
$\Lambda^- =\{ \la j : \epsilon _j =-\}$. Different
solutions will be parametrized by the sets
\define\gaaa#1{\gamma_1,\cdots,\gamma_{#1}}
$\gaaa {n-1}$ which will be specified later. The solutions look
as follows
$$
\align
&f^{\gaaa {n-1}}(\laaa {2n})_{\eee {2n}}=\\
&\prod_{\lambda^+ \in \Lambda^+,\lambda^- \in \Lambda^-}
(\lambda^+ - \lambda^-)^{-1}
\text{det}||\int_{\gamma_i}\omega_j(\tau|\Lambda^-|\Lambda^+)d\tau \ ||
_{(n-1)\times(n-1)}\tag 5
\endalign
$$
where $\omega_j$ are the following differentials on the HES
$w^2=P(\tau)\equiv $ \linebreak $\prod (\tau -\la i)$:
$$ \align &\omega_j(\tau|\Lambda^-|\Lambda^+)
=\frac{1}{\sqrt{P(\tau)}}\\ &\times
\bigl\{ \prod (\tau -\lambda^+)
\bigl [\frac{d}{d\tau}\frac { \prod (\tau -\lambda^-)}
{\tau^{n-j}} \bigr ]_0+
 \prod (\tau -\lambda^-)
\bigl [\frac{d}{d\tau}\frac { \prod (\tau -\lambda^+)}
{\tau^{n-j}} \bigr ]_0 \bigr \}
\endalign
$$
where $[\ ]_0$ means that only the polynomial part
of the expression in brackets is taken.
The differentials $\omega_j$ are of the second kind:
they have singularities at $\infty^{\pm}$, but their
residues at the infinities are equal to zero. It should be mentioned that
the singular part of the differential $\omega_j$ is
independent of the partition of $\Lambda$:
$$\omega_j(\tau|\Lambda^-|\Lambda^+)-
\omega_j(\tau|\Lambda^{\prime -}|\Lambda^{\prime +})=
\text{of the first kind} $$
First kind differentials in our case are of the type:
$\tau^j/\sqrt{P(\tau)},\ \ 0\le j \le n-2$.
The contours $\gaaa {n-1}$ are arbitrary cycles on the HES ( notice\linebreak
that its genus $g$ equals $n-1$ ). There are $2n-1$
evident cycles $c_i$ : $c_i$ contains two branching points
$(\la i,\la {i+1})$ $\ i=1,\cdots , 2n-1$, but only $2n-2$
of them are independent due to the relation
$$c_1+c_3+\cdots+c_{2n-1}\thicksim 0 $$
which implies that
$$ \sum_{i=1}^{n}\int_{c_{2i-1}} \omega_j=0,\ \ \ \forall j\tag 6 $$
The canonical $a$ and $b$ cycles can be introduced as follows:
$a_i=c_{2i-1},$\linebreak $ b_i=c_{2i}, \ i=1,\cdots, n-1$.
So, there are $C_{2g}^g$ independent solutions to (3). It can
be said that the solutions are parametrized by $\wedge^g TJ$
where $J$ is Jacobi variety of the HES. On the other hand
$TJ$ is exactly the place where $g$ independent local integrals
of classical periodical problem associated with $sl(2)$-type
integrable equations live. That is why the above formulae are
very suggestive from the point of view of the connection
between local and nonlocal ($\widehat{sl}(2)$) symmetries in
classical theory ( these problems are discussed in [12]).
It should be said also that the formulae (5) present a
variant of those given in [13]. They are more explicit due
to the peculiarity of $k=0$ case.

Now, let us turn to the problem of braiding. The equations (3) are
invariant under the permutation $\la i \leftrightarrow \la j$ and
simulteneous permutation of the  associated spaces. Let us denote
the operation of the analytical continuation
$\la i \leftrightarrow \la {i+1}$ and permutation of corresponding
spaces by $B_{i,i+1}$. Then we are supposed to have a formula of
the type:
\define\gaaaa#1{{\gamma}^{\prime}_1,\cdots,{\gamma}^{\prime}_{#1}}
$$ \align
&B_{i,i+1}f^{\gaaa {n-1}}(\laaa {2n})=\\
&=\sum_{\gaaaa {n-1}} C_{\gaaaa {n-1}}^{\gaaa {n-1}}
f^{\gaaaa {n-1}} (\laaa {2n})
\endalign
$$
where $C$ are some constants. It is well known nowadays
that the braiding is described by the finite-dimensional
quantum group $SL(2)_q$  [14], in our case $q=-1$. On the other
hand due to the explicit formulae (5) braiding allows different
interpretation being a consequence of recalculation of the
canonical bases of homology of HES after the automorphism
$\la i \leftrightarrow \la {i+1}$ is applied. So, we have
an amusing connection between $SL(2)_{-1}$ and very classical
mathematical problem.

2.{\bf Deformfed case ( Form factors for $su(2)$-invariant Thirring model).}

Let us consider the deformed case now. First, we introduce
the quantum R-matrix acting in $\text{C}^2 \otimes \text{C}^2$:
$$R_{1,2}=\frac{\beta-\pi i P_{1,2}}{\beta-\pi i}$$
where $P_{1,2}$ is permutation. This definition is different from
the definition of crossing-symmetrical S-matrix used in [3,4] by
certain factor. For the goals of the present paper we can
ignore this factor which is responsible only for the
multiplication of solutions of the deformed equations by
$\prod_{i,j}\zeta (\beta_i-\beta_j)$ with some rather complicated
function $\zeta$ [3]. Classical limit corresponds to
$\beta=\frac{2\pi i}{h}\lambda,\ \ h \to +i0 $. Evidently,
$$R_{1,2}(\beta)=I-hr_{1,2}(\lambda)+\Cal O (h^2)$$
in this limit.

For $k=0$ the deformed KZ equations [2] coincide essentially
with the Form Factors Equations [3]. But for our goals it is
more convenient to write them in the form used in [2]:
$$ \align
& F(\beta _1,\cdots ,\beta _j+2\pi i,\cdots ,\beta _{2n})\\
&= R_{2n,j} (\beta _{2n}-\beta _{j}-2\pi i)
\cdots R_{j+1,j} (\beta _{j+1}-\beta _{j}-2\pi i)\\
& \times R_{1,j} (\beta _{1}-\beta _{j})
\cdots R_{j-1,j} (\beta _{j-1}-\beta _{j})\\
&\times F(\beta _1,\cdots ,\beta _j,\cdots ,\beta _{2n})\tag 7
\endalign $$
where $F\in {(\text{C}^2)}^{\otimes 2n}$.
Formally, the equations (7) turn into (3) for the function
$$f(\laaa {2n})\simeq_{h\to +i0}C(h)F(\frac {2\pi i \lambda_1}{h},\cdots,
\frac {2\pi i \lambda_{2n}}{h}) \tag 8$$
where $C(h)$ is certain normalization constant which might be
needed. One point should be emphasized here. We did very
formal and potentially misleading computation. The real connection
between the solutions of deformed and undeformed equations can
be very complicated: the formula (8) can hardly present a real
limit, more probably it is asymptotical formula (which is
exactly the case in the situation under consideration), but
there may be even no mathematically clear correspondence
at all as it happens to be in $q$-deformed case for $|q|=1$
( the solutions of classical KZ equations with
trigonometrical r-matrix [15] do not describe even the asymptotics
of the form factors in Sine-Gordon model ).

The equations (7) allow explicit analytical solutions [3] which
were found even before then those of KZ equations. But before
proceeding further we have to introduce the notion of
quasiconstant. Evidently, the solutions of (7) can be multiplied
not only by constants, but also by arbitrary $2\pi i$-periodic
functions of $\beta_i$. However the multiplication by arbitrary
functions of the kind can disturb the braiding properties,
that is why we restrict ourselves with $2\pi i$-periodic
and symmetric functions. The ring of these functions is
generated by
$$ \align &z_i\equiv \sigma_i (\exp(\beta_1), \cdots ,\exp(\beta_{2n}))\\
& i=1,\cdots ,2n \endalign
$$
where $\sigma_i$ is elementary symmetrical polynomial of
degree $i$.

Now we shall describe the solutions. It should be mentioned
that only special ones were used for applications to form
factors which have to satisfy additional requirements. However
in implicit form the general solutions are contained in [3].

The usual base of the tensor product (4) is not useful for
our goals, so we start with introduction of another basis,
$w^{\eee {2n}} ( \beta_1 ,\cdots , \beta_{2n})$. The elements of
this basis satisfy the relations
$$ \align
&R_{i,i+1}(\beta_i - \beta_{i+1})
w^{\epsilon_1,\cdots,\epsilon_i,\epsilon_{i+1},\cdots,\epsilon_{2n}}
(\beta_{1},\cdots,\beta_i,\beta_{i+1},\cdots,\beta_{2n})\\
&=P_{i,i+1}w^{\epsilon_1,\cdots,\epsilon_{i+1},
\epsilon_{i},\cdots,\epsilon_{2n}}
(\beta_{1},\cdots,\beta_{i+1},\beta_{i},\cdots,\beta_{2n})
\endalign
$$
The vectors $w$ are explicitly described in [3]. For us it is essential
that
$$w^{\eee {2n}}(\beta_1,\cdots ,\beta_{2n}) \to e^{\eee {2n}}$$
as $\beta_k=\frac{2\pi i}{h}\la k,\ h\to 0$. So, in the limit $w$ reproduce
usual basis of $ (\text{C}^2)^{\otimes 2n}$. Let us denote the components
of
\define\beee {\beta_1,\cdots, \beta_{2n}}
\define\Beta{\text{B}}
$F(\beee)$ with respect to the basis $w$ by $F(\beee)_{\eee {2n}}$.
There are many solutions of (7) which are parametrized by sets
of integers
\define\kkk {k_1,\cdots,k_{n-1}}
$\{\kkk \}$ :\  $1 \le k_1<k_2<\cdots<k_{n-1}\le2n-1$. We again
consider only one-dimensional with respect to global $sl(2)$
solutions. The formulae for the solutions look as follows:
$$ \align &F^{\kkk}(\beee)_{\eee {2n}}
=\prod (\beta^+ -\beta^-)^{-1}\\
&\times \text{det}||\ (z_{2n})^{-1/2}z_{k_i}\int_{-\infty}^{\infty}
\Omega_j(\alpha|\Beta^-|\Beta^+)
e^{(n-k_i)\alpha}d\alpha\ ||_{(n-1)\times(n-1)}\tag 9
\endalign
$$
Here the same agreement about the partition of $\Beta=
\{\beee\}$ into $\Beta^-$ and $\Beta_+$ is done as in the
formula (5). The functions $\Omega_j$ (the deformed analogs
of $\omega_j$ ) are as follows
\define\GGG#1{\Gamma(\frac{1}{4}+\frac{#1}{2\pi i})
\Gamma(\frac{1}{4}-\frac{#1}{2\pi i})}
$$ \align
&\Omega_j(\alpha|\Beta^-|\Beta^+)=\prod_{j=1}^{2n}
\GGG {\alpha-\beta_j}\\
& \bigl\{ \prod(\alpha-\beta^++\frac{\pi i}{2})Q_j(\alpha|\Beta^-)
+ \prod(\alpha-\beta^--\frac{\pi i}{2})Q_j(\alpha-\pi i|\Beta^+-\pi i)
\bigr\}
\endalign
$$
where $\Beta^{+}-\pi i=\{\beta^{+}-\pi i\}$,
$$Q_j(\alpha|\Beta)=\sum_{l=0}^j \bigl[ (\alpha+\frac{\pi i}{2})^l-
 (\alpha-\frac{\pi i}{2})^l\bigr]\sigma_{j-l}(\Beta)$$
where $\sigma_m$ are elementary symmetrical polynomials of degree $m$.
The limitations on $\kkk$ are due to the requirement of the
convergency of the integrals, notice that
$$ \GGG {\alpha}\thicksim_{\alpha\to\pm\infty}
|\alpha|^{-1/2}\text{exp}(-\frac{1}{2}|\alpha|)\tag 10$$
The quasiconstansts $z_{k_i}$ in the formula (9) are introduced
for the sake of better classical limit.

Just like in undeformed case the real number of independent
solutions is less than the number of partitions $\kkk$ due to
the following remarkable identity [3]:
$$ \align
& \sum_{i=1}^n z_{2i-1} \int_{-\infty}^{\infty}\Omega_j(\alpha|\Beta^-|\Beta^+)
e^{(n-(2i-1))\alpha}d\alpha=0\\
&\forall j,\Beta^-,\Beta^+\tag 11
\endalign
$$
This identity can be considered as exact deformation of (6).
After the relation (11) is imposed we find $C_{2n-2}^{n-1}$
independent solutions of (7) which is exactly the same as in
undeformed case. The physically important solutions of (7)
which describe the energy-momentum tensor for $su(2)$-
invariant Thirring model correspond to the following special
choice of $\kkk$:\ $k_i=2i-1,\ \ i=1,\cdots,n-1$.

Let us now turn to the problem of braiding. Notice first that
the functions (9) are meromorphic functions of all their
arguments with essential singularities at infinity. This
property differs them from the solutions of undeformed
equations which have branching points at $\la i=\la j$.
So, it is very hard for the solutions of deformed equations
being a meromorphic functions to have nontrivial braiding.
Let us consider that more formally.
The equations (7) are invariant under the permutation
$\beta_i \leftrightarrow \beta_j$ with simulteneous
multiplication of $F$ by $R_{i+1,i}(\beta_{i+1}-\beta_i)$.
Let us denote the operation of analytical continuation
$\beta_i \leftrightarrow \beta_j$ together with
multiplication of $F$ by $R_{i+1,i}(\beta_{i+1}-\beta_i)$ by
$B_{i,i+1}$. We are supposed to have a relation of the kind:
\define\kkkk{k_1^{\prime},\cdots,k_{n-1}^{\prime}}
$$\align
&B_{i,i+1}F^{\kkk}(\beta_{1},\cdots,\beta_{2n})\\
&=\sum_{\kkkk} C_{\kkkk}^{\kkk}F^{\kkkk}
(\beta_{1},\cdots,\beta_{2n})
\endalign $$
But it is quite clear from (9) that
$$ C_{\kkkk}^{\kkk}=\delta_{\kkkk}^{\kkk}$$
so, there is no mixing different solutions
braiding for the equations (7), and we could
impose the symmetry condition from the very beginning as it
is done in [3]. This remarkable difference between the solutions of
deformed and undeformed equations makes us to consider more carefully
the classical limit $ \beta=\frac{2\pi i}{h},\ \ h\to+i0$.

3.{\bf Classical limit.}

To start this section we shall first formulate the results of it.
There is one-to-one correspondence between the solutions of
deformed and undeformed equations. But the correspondence is
asymptotical: every particular solution of (3) describes the asymptotics
of one solution of (7) when $h\to+i0$. The asymptotical character
of the correspondence explains the differences of the
properties of the solutions.

Before proceeding further let us illustrate the last
statement with simple example. Consider the function;
$$f(x)=\text{ch}(x/2)\Gamma(\alpha+\frac{x}{2\pi i})
\Gamma(\alpha-\frac{x}{2\pi i})$$
which satisfies the functional equation:
$$f(x+2\pi i)=\frac{2\pi i\alpha+x}{2\pi i(1-\alpha)+x}f(x)\tag 12$$
Evidently, $f(x)$ is meromorphic function of $x$ with
essential singularity at the infinity. Suppose now
$x \in\text{R}$ and consider the asymptotics $x=\Lambda y,\
\Lambda \to \infty$ :
$$ f(\Lambda y)\thicksim\Lambda^{\kappa}\ (|y|)^{\kappa},\ \ \kappa=
2\alpha -1$$
as a consequence of Stirlling formula. So, the asymptotics is
not analytical in $y$ ( contains $|y|$ ). Consider now $y>0$ ,
calculate the asymptotics, and then continue it analytically
the result being:
$$f_0(y)=\Lambda^{\kappa}\ (y)^{\kappa}$$
which has branching point at $y=0$ and is not even but
satisfies the formal limit of the equation (12):
$$\frac{d}{dy}f_0(y)=\frac{\kappa}{y}f_0(y)$$
This is the type of phenomena we deal with. The mechanism
responsible for them is the concentration of infinite
sequences of poles into cuts.

Let us calculate now the asymptotics of (9) for
$\beta_j=\frac{2\pi i}{h}\lambda_j,\ \ h\to+i0$. We have first
to order $\la j$ in certain way, let it be $\la 1<\la 2<\cdots
<\la {2n}$. The main problem is in estimating of the integrals
in the determinant. The way of doing that is explained in [4,16].
Let us take one of the integrals:
$$(z_{2n})^{-1/2}z_{k_i}
\int_{-\infty}^{\infty}\Omega_j(\alpha|\Beta^-|\Beta^+)
e^{(n-k_i)\alpha}d\alpha $$.
We have rescaled $\beta_j$ as $\frac{2\pi i}{h}\lambda_j$, let us
also rescale $\alpha$ as $\frac{2\pi i}{h}\tau$. The function
$\GGG {\alpha-\beta_j}$ behaves as $h\to+i0$ as follows:
$$\GGG {\alpha-\beta_j}\thicksim |h|^{1/2}(|\tau-\la j|)^{-1/2}
\text{exp}(-\frac{1}{2|h|}|\tau-\la j|)$$
Hence, we have very rapid exponential behaviour, so, let us concentrate
on it for a moment neglecting the powers.
\define\seg#1{\lambda_{#1}<\tau<\lambda_{#1+1}}
Inside the segment $\seg m$ the exponential behaviour of the
integrand is the following
$$\text{exp}({1\over |h|}((2n-m-k_i)\tau+1/2\sum_{k=1}^{m}\la k-
1/2\sum_{k=m+1}^{2n}\la k))$$
which means that for $2n-m<k_i$ the integral is estimated due to
the integration by parts by the value of the integrand at the
left, while for $2n-m>k_i$ -- at the right end of the segment.
The segment $\seg {2n-k_i}$ is of special importance: here is the
only place where the integration survives. Certainly these reasonings
should be supported by study of the power contributions. Due to
(10) there are singularities at $\tau=\la i$, but fortunetly they
are integrable, so, they only change $h$ in the integration
by parts estimation to $h^{1/2}$. Combining all that, and performing
exact calculation we draw the following conclusion:
the main contribution to the asymptotics is provided by the
integral over $\seg {2n-k_i}$, the contributions from $\seg m,\
m>2n-k_i+1,m<2n-k_i-1$ are exponentially small, while the integrals
over $\seg {2n-k_i\pm1}$ are of the order $h^{1/2}$ with respect to
the leading one.
So, the final result is the following:
$$ \align
&(z_{2n})^{-1/2}z_{k_i}\int_{-\infty}^{\infty}
\Omega_j(\alpha|\Beta^-|\Beta^+)
e^{(n-k_i)\alpha}d\alpha\\
&=h^{-(j-1)}\bigl[\int_{\la {2n-k_i}}^{\la {2n-k_i+1}}
\omega_j(\tau|\Lambda^-|\Lambda^+)d\tau+\Cal{O}(h^{1/2})\bigr]
\endalign $$
We used the fact that the polynomial part of $\Omega_j$
evidently tends to that of $\omega_j$ in the limit.
Evidently,
$$
\int_{\la {m}}^{\la {m+1}}\omega_j(\tau|\Lambda^-|\Lambda^+)d\tau
=1/2\int_{c_ {m}}\omega_j(\tau|\Lambda^-|\Lambda^+)d\tau
$$
where $c_i$ are the cycles on HES introduced in Section 1.

Thus,
$$\align &F^{\kkk}(\beee)\\
&=(1/2)^{n-1}h^{\frac{(n-1)(n-2)}{2}}\bigl[f^{\gaaa {n-1}}(\laaa {2n})
+\Cal{O}(h^{1/2})\bigr] \tag 13\endalign $$
where $\gamma_i=c_{2n-k_i}$ . The asymptotical formula (13) proves the
statements formulated at the beginning of this section.
Actually, the asymptotical correspondence is so explicit that
it makes us to suspect certain structure of ``quantized'' HES
underlying it.
The differentials
$$ \alpha^j\prod_{j=1}^{2n}\GGG {\alpha-\beta_j}d\alpha,\ \ 0\le j \le n-2 $$
play the role of first kind (regular holomorphic) differentials.
The differentials
$$\Omega_j (\alpha|\text{B}^-|\text{B}^+)d\alpha,\ \ 1\le j\le n-1$$
for any particular partition into $\text{B}^-$ and $\text{B}^+$
can be taken for the base of holomorphic differentials singular at
infinity. The singular part of the differentials is independent on
the partition:
$$\Omega_j (\alpha|\text{B}^-|\text{B}^+)d\alpha
-\Omega_j (\alpha|\text{B}^{\prime -}|\text{B}^{\prime +})d\alpha =
\text{``first kind''}$$
There is the following correspondence between the
integrals:
$$\int_{c_{2n-l}}\leftrightarrow z_{2n}^{-1/2}z_l
 \int_{-\infty}^{+\infty}\text{exp}((n-l)\alpha)$$
The relation (11) is very important from that point of view.

{\bf Conclusion.}

Let us formulate several challenging problems arising
in the connection with the matters considered in the
paper.

1. As it has been already mentioned there should be
a connection between $k=0$ KZ equations and finite-gap
solutions of classical integrable equations. This
connection should imply beautiful relation between
local integrals and nonlocal ones. The latter are associated to
monodromy being described in our case by $\widehat{sl}(2)$
while the first ones are associated with flows on the
Jacobi variety of the HES. This is exactly the kind
of things we deal with for $k=0$ KZ equations.

2. The solutions of $k=0$ KZ equations provide enough
amount of data to describe the HES: all the periods
of holomorphic differentials can be extracted from them.
It would be interesting to understand more about this
connection of algebraic and analytic objects.

3. When the connection between the loop algebra and
HES is understood it should be possible to understand the
``quantization'' of HES mentioned above, because algebraic
mechanism responsible for that is the deformation of
the loop algebra into the double of Yangian.

4. The problem of generalization of our results
is also important. HES arise in connection with $\widehat{sl}(2)$.
In the paper [17] (see also[16]) the form factors of $su(N)$-invariant
Thirring model were calculated which satisfy from modern point
of view the deformed $sl(N),k=0$ KZ equations. The solutions are
given in terms of integrals which can be regarded as deformations
of periods of differentials on the surfaces with the branching
points of order $N$. Is it possible to deal with more general
type of surfaces and what kind of algebraic structure corresponds
to them?

\newpage

{\bf References}

\def\no#1#2\par{\item{#1.}#2\par}
\def\jr#1{{\nineit#1}}

\def\vl#1{{\ninebf#1}}

\no 1
A.G. Knizhnik, A.B. Zamolodchikov,
\jr{Nucl. Phys. B}
\vl {297}
(1984) 83

\no 2
I.B. Frenkel, N.Yu. Reshetikhin,
Holonomic Systems of of q-Difference Equations,
Yale prepint (1991)

\no 3
F.A. Smirnov, \jr{Journal of Physics}
\vl 19A
(1986) L575

A.N. Kirillov, F.A. Smirnov, \jr{Phys. Lett.}
\vl 198B
(1987) 506

F.A. Smirnov,
in Introduction to Quantum Groups and Integrable
Massive Models of Quantum Field Theory, \jr{Nankai Lectures on
Mathematical Physics }Mo-Lin Ge and Bao-Heng Zhao eds.
World Scientific (1990)

\no 4
F.A. Smirnov,
Dynamical symmetries of massive integrable models,
RIMS preprints RIMS-772,RIMS-838 (1991)

\no 5
A.Yu. Alekseev, L.D. Faddeev, M.A. Semenov-Tian-Shansky,
Hidden Quantum Groups inside Kac-Moody Algebras,
LOMI preprint,E-5-91
(1991) 20p.

\no 6
A.B. Zamolodchikov,
\jr{Int. Journ. of Mod. Phys.}
\vl{A4}
(1989) 4235; in
\jr{Adv. Studies in Pure Math.}
\vl{19}
(1989) 641

\no 7
L.D. Faddeev,
{}From Integrable Models to Quantum Groups.
\jr{In: Fields and Particles,} H.Mitter and W.Schweiger eds.
Springer
(1991) 89

\no 8
V.G. Drinfeld,
\jr{Sov. Math. Dokl.}
\vl{32}
(1985) 254.

V.G. Drinfeld,
Quantum Groups,
\jr{Proceedings of the International Congress}

\jr{of Mathematicians,}
Berkeley, CA, (1987) 798

\no 9
M. L\"uscher,
\jr{Nucl. Phys. B}
\vl{135}
(1978) 1

\no {10}
D. Bernard, \jr{CMP}
\vl{137}
(1991), 191

\no {11}
A. LeClair, F.A. Smirnov, Infinite quantum group symmetry of
fields in massive $2D$ quantum field theory, Cornell preprint
CLNS 91-1056 (1991)

\no {12}
O. Babelon, D. Bernard,
Dressing Symmetries,
Preprint SPhT-91-166,\linebreak LPTHE-91-56,
(1991)

\no {13}
P. Christe, R. Flume,
\jr{Nucl. Phys.}
\vl{B282}
(1987) 466.

V.V. Schechtman, A.N. Varchenko,
Integral Representations of $N$-point

Conformal Correlations in the
$WZW$ Model,
Max-Plank-Institute preprint

(1987)

\no {14}
A. Tsuchiya, Y. Kanie,
In: Conformal Field Theory and
Solvable Lattice Models,
\jr{Adv. Stud. Pure. Math.}
\vl{16}
(1988) 297.

\no {15}
I. Cherednik,
Integral Solutions of Trigonometric KZ Equations and KM algebras,
Preprint RIMS-699
(1990)

\no {16}
F.A. Smirnov,
Form Factors in Completely Integrable Models of Quantum
Field Theory,
\jr{to be published in World Scientific}

\no {17}
F.A. Smirnov,
Form Factors for SU(N)-Chiral Gross-Neveu Models.
 Preprint \linebreak LOMI E-1-88,
(1988)

\enddocument